%% file: ARXIV.tex
\documentclass[journal,onecolumn]{IEEEtran}

\input{preamble_arxiv.tex}

\begin{document}
\title{On the connection between the ABS perturbation methodology and differential privacy\footnote{Views expressed in this paper are those of the authors and do not necessarily represent those of the ABS. Where quoted or used, they should be attributed clearly to the authors. The work of P.~Sadeghi was supported by the Australian Research Council Future Fellowship FT19100429.}}

\author{\IEEEauthorblockN{Parastoo Sadeghi$^{*}$ and Chien-Hung Chien$^{\dag}$}\\
\IEEEauthorblockA{$^{*}$School of Engineering and IT, University of New South Wales, Canberra, Australia}\\
\IEEEauthorblockA{$^{\dag}$Methodology Division, Australian Bureau of Statistics, Canberra, Australia}\\
Emails: \url{p.sadeghi@unsw.edu.au;joseph.chien@abs.gov.au}}

\maketitle

\begin{abstract}
	\noindent 
	This paper explores analytical connections between the perturbation methodology of the Australian Bureau of Statistics (ABS) and the differential privacy (DP) framework. We consider a single static counting query function and find the analytical form of the perturbation distribution with symmetric support for the ABS perturbation methodology. We then analytically measure the DP parameters, namely the $(\eps, \delta)$ pair, for the ABS perturbation methodology under this setting. The results and insights obtained about the behaviour of $(\eps, \delta)$ with respect to the perturbation support and variance are used to judiciously select the variance of the perturbation distribution to give a good $\delta$ in the DP framework for a given desired $\eps$ and perturbation support. Finally, we propose a simple sampling scheme to implement the perturbation probability matrix in the ABS Cellkey method. The post sampling $(\eps, \delta)$ pair is numerically analysed as a function of the Cellkey size. It is shown that the best results are obtained for a larger Cellkey size, because the $(\eps, \delta)$ pair post-sampling measures remain almost identical when we compare sampling and theoretical results.  	
\end{abstract}

\section{INTRODUCTION}

The Australian Bureau of Statistics (ABS) is committed to improving access to ABS statistics, while continuing to ensure privacy and confidentiality are maintained \cite{ABS2022}. The emergence of differential privacy (DP) methods provides opportunities to better quantify the trade-off between statistical utility and confidentiality protection in statistical outputs. As a result, the ABS is continuing to explore the opportunities offered by DP. This research builds on \cite{Fraser2005, Marley2011} and \cite{Bailie2019} and seeks to enhance the perturbation methodology in the ABS TableBuilder through the lens of DP. ABS perturbation methodology has two components – an entropy maximisation method for generating the perturbation probability transition matrix (or the perturbation table) and a cell key method to ensure consistent protections for statistical outputs \cite{Fraser2005, Marley2011}. At a high level, this work improves both components by first proposing an approach to incorporate the DP framework while creating the perturbation table and then developing a sampling scheme to make full use of the perturbation table using a memory-efficient lookup table. Overall, this work offers tools and insights for analytical quantification of DP measures for the ABS perturbation methodology and improves its implementation efficiency. To the best of our knowledge, this is the first attempt to  quantify the connection between ABS perturbation methodology design parameters and DP metrics in an analytical and provable format. While we consider a specific case with a single counting query and a symmetric perturbation support, the methodology and insights have the potential to be extended to more advanced and complex cases with multiple counting queries or asymmetric perturbation support. 

More specifically, our contributions include: (1) introducing a method to analytically quantify the $\epsilon$ and $\delta$ DP parameters in the ABS perturbation methodology for a single counting query with symmetric perturbation support; (2) developing an approach to incorporate the $\epsilon$ DP parameter and the symmetric support of the distribution into the entropy maximisation process; (3) showing the importance of carefully choosing the variance parameter in the method proposed by \cite{Fraser2005,Marley2011} with respect to the DP parameters; and (4) proposing a sampling scheme to ensure the proposed method can be efficiently integrated with the cell key approach to improve ABS perturbation methodology and quantifying the ($\epsilon$,$\delta$)-DP parameters post sampling.

The paper is structured as follows. Section \ref{sec:preliminary} provides the key notation and describes the entropy maximisation proposed by \cite{Fraser2005, Marley2011}.  Section \ref{sec:main} discusses the proposed analytical entropy maximisation approach to incorporate ($\epsilon$,$\delta$)-DP parameters for noise distributions with symmetric supports. We propose an approach to quantise and sample the probability mass function (pmf) with a simple lookup table in Section \ref{sec:cellkey}. We show the importance of increasing the size of the row index look up in Section \ref{sec:sampling}. Finally, we provide a conclusion and propose future research directions in Section \ref{sec:conclusion}.

\section{SYSTEM MODEL AND PRELIMINARIES} \label{sec:preliminary}

This section provides the notation conventions used throughout the paper. The set $\{a, \cdots, b\}$ for some $a,b \in \mathbb{Z}$, $a \leq b$ is compactly represented as $[a,b]$. 

We consider a single counting query function $q$ from a dataset $x \in \X$. Assume the true count is $q(x) = n$. In order to enhance the privacy of individuals in the dataset, a discrete-valued independent random variable $Z$ with alphabet $\Z$ and probability mass function (pmf) $p_Z$ is added to the true count to give the random query response
\begin{align}\label{eq:mechanism}
M(x) = q(x) + Z.
\end{align}
For brevity, we may simply refer to $Z$ as noise. The parameters of the noise pmf are assumed to be independent of the dataset $x$. The probability mass of noise at $z\in \Z$ is denoted by $p_Z(Z=z)$. We may use the short-hand notation $p(z)$ where the context is clear.

References \cite{Fraser2005, Marley2011} show that given the above model and assumptions, the ABS TableBuilder aims to maximise statistical confusion induced by noise, measured by the Shannon entropy. It performs the following constrained optimisation to derive the noise parameters
\begin{align}\label{ABS:optGen}
&\max_{P_Z} H(Z) = \max_{P_Z} \sum_{z\in \Z} p(z)\log\frac{1}{p(z)},\\
&\qquad \text{s.t.}
\begin{cases}
\qquad \E[Z] = 0, &\qquad \text{zero bias,}\\ \qquad\E[Z^2] \leq V, &\qquad \text{variance constraint,}\\\qquad \sum_{z \in \Z} p(z) = 1, &\qquad \text{valid pmf,}
\\\qquad p(z) \geq 0, \quad \forall z \in \Z,&\qquad \text{valid pmf.}
\end{cases}
\end{align}
We use the natural logarithm in this paper.

In this paper, we are interested in analytically characterising and improving the differential privacy measure of the TableBuilder. To this end, we recall the definition of differential privacy from \cite{dwork2006calibrating, dwork2014algorithmic}. Throughout this paper, $\eps \in \mathbb{R}^+$.

\begin{definition}(Approximate Differential Privacy) A randomised mechanism $M: \X \to \Y$ is said to satisfy $(\eps, \delta)$-differential privacy or $(\eps, \delta)$-DP for short, if for all datasets $x,x'\in \X$ differing on a single element and all events $E \subset \Y$, we have 
$$\Prob[M(x) \in E ]\leq e^\eps \Prob[M(x') \in E] +\delta.$$ If $\delta = 0$, we obtain pure or just $\eps$-DP. If $0< \delta \leq 1$, we obtain approximate $(\eps, \delta)$-DP. 
\end{definition}

\section{MAIN RESULTS}\label{sec:main}

\subsection{Analytical Distribution of the Symmetric TableBuilder Noise}\label{sec:pZ}

The noise range $\Z$ in the TableBuilder method is general and can be any subset of the integers $\mathbb{Z}$. However, to analytically characterise and optimise the differential privacy performance of the TableBuilder, we focus on the special symmetric case where $\Z = [-D,D]$ for some $D \in \N$.  In order for the random query output $M(x)$ to also remain non-negative, we further assume the true count satisfies $q(x) = n \geq D$. With these assumptions, we specialise the TableBuilder optimisation problem in \eqref{ABS:optGen} as
\begin{align}\label{ABS:optD}
&\max_{P_Z} H(Z) = \max_{P_Z} \sum_{Z = -D}^{D} p(z)\log\frac{1}{p(z)},\\
&\qquad \text{s.t.}
\begin{cases}
\qquad \sum_{Z = -D}^{D} zp(z) = 0, &\qquad \text{zero bias,}\\ \qquad\mathbb \sum_{Z = -D}^{D} z^2p(z) \leq V, &\qquad \text{variance constraint,}\\\qquad \sum_{z=-D}^D p(z) = 1, &\qquad \text{valid pmf,}
\\\qquad p(z) \geq 0,\quad z \in [-D,D]&\qquad \text{valid pmf.}
\end{cases}
\end{align}
Taking the derivative of the Lagrangian function for this problem and after some manipulations, the optimal distribution $p(z)$ is of the form
\begin{align}\label{eq:pz11}
p(z) = C e^{-\gamma z^2}, \qquad z \in [-D,D],
\end{align}
where $C$ is the normalisation constant and is given as
\begin{align}\label{eq:C1}
\sum_{z = -D}^{D} C e^{-\gamma z^2} = 1 \qquad \Rightarrow \qquad C = \frac{1}{2\sum_{z=1}^D e^{-\gamma z^2}+1}.  
\end{align} 
The parameter $\gamma$ is chosen to  satisfy the variance constraint
\begin{align}\label{eq:V1}
\sum_{z=-D}^D z^2 C e^{-\gamma z^2}  = 2C \sum_{z=1}^D z^2 e^{-\gamma z^2} = V.
\end{align}
Combining \eqref{eq:C1}  and \eqref{eq:V1} together, we get
\begin{align}\label{eq:gamma1}
&2 \sum_{z=1}^D z^2 e^{-\gamma z^2} = V (2\sum_{z=1}^D e^{-\gamma z^2}+1) \qquad \Rightarrow \qquad \sum_{z=1}^D (2z^2-2V)e^{-\gamma z^2} - V = 0.
\end{align}
Let us denote $x:= e^{-\gamma}  > 0$. To find the pmf of noise, we need to numerically solve the following polynomial equation of degree $D^2$ in $x$:
\begin{align}\label{eq:gammax1}
f(x) := \sum_{z=1}^D (2z^2-2V)x^{ z^2} - V = 0.
\end{align}
This equation has sparse nonzero coefficients at square degrees $D^2, (D-1)^2, \cdots, 9, 4, 1, 0$.

It is desirable for $p_Z$ to have its highest probability at $Z = 0$ (corresponding to the truthful count $M(x)=q(x) = n$ having the highest likelihood in the response). That is, we wish to have $0 < e^{-\gamma} < 1 \Leftrightarrow \gamma > 0$. This means the polynomial $f(x)$ must have a root between $0$ and $1$.  Note that $f(0) = -V < 0$. Note also that
\begin{align*}
f(1) &= \sum_{z=1}^D (2z^2-2V)- V = \frac{D(D+1)(2D+1)}{3}-(2D+1)V
=(2D+1)\left(\frac{D(D+1)}{3}-V\right).
\end{align*}
Therefore, if $f(1) > 0$, then $f(x)$ is guaranteed to have a root between 0 and 1. For $f(1) > 0$, we have the following proposition.
\begin{proposition}\label{prop:var}
For the TableBuilder pmf with symmetric support $\Z = [-D,D]$ to be a decreasing function of $|z|$, its variance $V$ should satisfy
\begin{align}
0 < V < \frac{D(D+1)}{3}.
\end{align}
\end{proposition}
This bound on variance $V$ is consistent with the fact that among all probability mass functions over the support $[-D,D]$, the uniform distribution has the maximum entropy $H(Z) = \log(2D+1)$, zero bias, and variance  $\frac{D(D+1)}{3}.$ In the rest of Section \ref{sec:main}, we will impose the constraint in Proposition \ref{prop:var}.

\subsection{Differential Privacy Parameters of the ABS TableBuilder Method}\label{sec:tablebuilderDP}

We take a first-principles approach to computing the $(\eps, \delta)$-DP parameters of the TableBuilder mechanism. Our approach is similar in spirit to the one introduced in \cite{balle18a} for the continuous Gaussian mechanism. However, the derivation of $\delta$ and optimisation of the TableBuilder noise pmf are very different and a main novelty of this paper.

Throughout this subsection, we assume the TableBuilder noise support $[-D,D]$ and noise variance $V$ are given. Recall that the variance $V$ determines $\gamma$ in \eqref{eq:pz11}, which is found via solving \eqref{eq:gammax1}. In summary, the TableBuilder noise pmf $P_Z$ in \eqref{eq:pz11} is parameterised by $D$ and $\gamma$. 

In this subsection, we characterise $\delta$ as a function of $\eps$ for given TableBuilder noise parameters $D$ and $\gamma$. To make these dependencies clear, we denote it as $\delta_{\gamma, D}(\eps)$. In the next subsections, we take the analysis one step further, where we will study and optimise the effect of the TableBuilder noise parameters, $\gamma$ and $D$, on $\delta$.

The principle for characterising $\delta_{\gamma, D}(\eps)$ is as follows 
\begin{align}\label{eq:deltazJ}
\delta_{\gamma, D}(\eps) &= \sum_{z \in E^*} p(z) - e^{\eps}p(z-1),\\
E^*:= E^*_{\gamma, D}(\eps)&\triangleq \{z: z \in [-D,D+1], \frac{p(z)}{p(z-1)}>e^\eps\},
\end{align}
where the set $E^{*}$ captures outputs $y$ for which the privacy loss function exceeds the desired privacy level $e^{\eps}$.\footnote{We interpret $0/0 = 0$. This happens when both pmf values $p(z)$ and $p(z-1)$ are zero and hence do not contribute to the privacy loss.} We first note that $ z = -D$ always belongs to  $E^*$ regardless of $\eps$ and $\gamma$. This is because $p(Z=-D) \neq 0$ and  $p(Z=-D-1) = 0$ resulting in the privacy loss ratio becoming $\infty$. Therefore, a simple lower bound on $\delta$ is the noise pmf value at $Z = -D$. That is,
\begin{align}\label{eq:deltaLB}
\delta_{\gamma, D}(\eps) \geq p_Z(Z=-D) = C e^{-\gamma D^2}, \qquad \forall \eps > 0.
\end{align}
In order to fully characterise the set $E^*$, let us expand and simplify it as
 \begin{align}\label{eq:Estarsimp}
E^*_{\gamma, D}(\eps) &\triangleq \left\{z: z \in [-D,D+1], \quad \frac{p(z)}{p(z-1)} = \frac{e^{-\gamma z^2}}{e^{-\gamma (z-1)^2}}> e^\eps\right\}\\
& = \{z: z \in [-D,D+1], \quad e^{-2\gamma z + \gamma}> e^\eps\}\\
& = \{z: z \in [-D,D+1], \quad -2\gamma z + \gamma>\eps\}\\
& =  \left\{z: z \in [-D,D+1], \quad z < 0.5-\frac{\eps}{2\gamma}\right\}.
\end{align}

Under the constraint detailed in Section \ref{sec:pZ} that $V < \frac{D(D+1)}{3}$, we will have that  $\gamma > 0$ and hence, $z \geq 1$ cannot belong to $E^*$. Therefore, it suffices to determine whether each $ z \in [-D+1:0]$ belongs to $E^*$ or not. Let us define
 \begin{align}\label{eq:Estar1}
F^*:=F^{*}_{\gamma, D}(\eps) &\triangleq  \left\{z: z \in [-D+1:0], \quad z < 0.5-\frac{\eps}{2\gamma}\right\}.
\end{align}
Denote $z^* := \lfloor 0.5-\frac{\eps}{2\gamma}\rfloor$. We consider two cases:
\begin{enumerate}
\item $0 < \eps < \gamma\quad \Rightarrow \quad 0 <  \frac{\eps}{2\gamma} < 0.5 \quad \Rightarrow \quad 0 < 0.5-\frac{\eps}{2\gamma} < 0.5$. Therefore, $z^* = 0$ and $F^* = \{-D+1, \cdots, 0\}$
\item $ \eps > \gamma > 0 \quad \Rightarrow \quad \frac{\eps}{2\gamma}  > 0.5 \quad \Rightarrow \quad 0.5-\frac{\eps}{2\gamma} <0$ and hence $z^* < 0$. Within this case, there are two sub-cases:
\begin{enumerate}
\item  If $z^* \leq -D$,  then $F^* = \emptyset$. This means $E^* = \{-D\}$.
\item  If $ -D+1\leq z^* < 0$,  then $F^* = [-D+1, z^*] \neq \emptyset$.
\end{enumerate}
\end{enumerate}
Therefore, the set $E^*$ can be compactly written as
 \begin{align}\label{eq:Estar22}
E^{*}_{\gamma, D}(\eps) = \{-D\}\cup F^{*}_{\gamma, D}(\eps) = [-D, \max\{-D, z^* \}].
\end{align}
In summary, we analytically characterise $\delta_{\gamma, D}(\eps)$ in the following proposition. 

\begin{proposition}\label{prop:delta}
Consider the mechanism in \eqref{eq:mechanism} for the single counting query $q$. The TableBuilder mechanism with noise pmf given in \eqref{eq:pz11}-\eqref{eq:V1} and variance $V$ satisfying Proposition \ref{prop:var} achieves $(\eps,\delta)$-DP such that
 \begin{align}\label{eq:deltafinal1}
\delta = \delta_{\gamma, D}(\eps) &= \begin{cases}
Ce^{-\gamma D^2}, &\quad  \lfloor 0.5-\frac{\eps}{2\gamma}\rfloor \leq -D,\\
Ce^{-\gamma D^2}+C\sum_{z =-D+1}^{\lfloor 0.5-\frac{\eps}{2\gamma}\rfloor}  (e^{-\gamma z^2}- e^{\eps} e^{-\gamma (z-1)^2}), &\quad  -D < {\lfloor 0.5-\frac{\eps}{2\gamma}\rfloor} \leq 0,
\end{cases}
\end{align}
where $\gamma$ is determined by $V$ via solving the polynomial equation in \eqref{eq:gammax1}.
\end{proposition}

In Fig. \ref{fig:deltas}, we present evaluation of $\delta_{\gamma, D}(\eps)$ according to \eqref{eq:deltafinal1} for four possible combinations of $D = 11$ and $D=15$ with $\gamma = 0.125$ and $\gamma \approx 0.0498$ (corresponding to two variances $V = 4$ and $V = 10$, respectively). They are divided into Fig. \ref{fig:deltasA} and \ref{fig:deltasB} for different values of $D$. There are a number of important observations that can be made from the two figures. Broadly speaking, when $\gamma$ is fixed, increasing the noise support span $D$ will decrease $\delta$. However, the impact of $\gamma$ on $\delta$ and its interactions with $\eps$ is complex. \cite{Fraser2005} and \cite{Marley2011} rely on variance $V$ and support $D$ to design the noise and there are no specific relationships between $V$ (or $\gamma$), $\delta$ and $\eps$. Therefore when we attempt introducing $\eps$ parameter to calculate $\delta$ using \eqref{eq:deltafinal1}, we can observe increasing $\eps$ will hit a point where $\delta$ is not decreasing.  This is because if the relation of $\eps$ with $D$ and $\gamma$ is such that $\lfloor 0.5-\frac{\eps}{2\gamma}\rfloor \leq -D$, we will have a fixed $\delta = Ce^{-\gamma D^2}$ regardless of how much larger $\eps$ gets. Figure \ref{fig:deltasA} show that $\delta$ hits a plateau after reaching a certain point in $\eps$. The \cite{Fraser2005, Marley2011} preserve better confidentiality-utility trade off before hitting this plateau where $\delta$ is lower for a higher variance $V$. However, the plateau in $\delta$ prevents it from continuing this trend. Overall, it is useful to note that a careful choice of parameters for the TableBuilder noise is needed to ensure a desired outcome. We will discuss this topic in greater detail in subsection $\ref{sec:optvar1}$ and $\ref{sec:designguide}$. 

\FigOneTBNoise{Plots of $\delta_{\gamma, D}(\eps)$ using \eqref{eq:deltafinal1} for $D = 11$ and two values of $\gamma = 0.125$ and $\gamma \approx 0.0498$ corresponding to two variances $V = 4$ and $V = 10$, respectively.}{Plots of $\delta_{\gamma, D}(\eps)$ using \eqref{eq:deltafinal1} for $D = 15$ and the same variances $V=4$ and $V=10$.}{Plots of $\delta_{\gamma, D}(\eps)$ for various TableBuilder noise parameters $D$ and $\gamma$.}

\subsection{Selection of TableBuilder Parameters under $(\eps, \delta)$-DP Framework}\label{sec:optvar1} 

In the previous subsection, we derived the $(\eps, \delta)$-DP parameters of the Tablebuilder mechanism. The derivation technique takes the TableBuilder $\gamma$ and $D$ as input parameters and determines what $\delta$ is achievable as a function of $\eps$. We observed that for a fixed $\gamma$, there comes a threshold in $\eps$ beyond which increasing $\eps$ does not decrease $\delta$. We attributed this plateauing phenomenon to the existence of the first case for $\delta_{\gamma, D}(\eps)$ in \eqref{eq:deltafinal1} in Proposition \ref{prop:delta}. Even as we increase $\eps$, we observed that $\delta_{\gamma, D}(\eps)$ is bounded away from zero by $Ce^{-\gamma D^2}$. In this subsection, the core idea is to judiciously select $\gamma$ (or variance) as a function of $\eps$ to avoid a plateau in $\delta$. 

As we know from the first case in \eqref{eq:deltafinal1}, making $z^*$ smaller than $-D$ by increasing $\eps$ does not result in a reduction of $\delta$. Therefore, we propose to choose $\gamma(\eps)$ such that $z^* = \lfloor 0.5-\frac{\eps}{2\gamma}\rfloor = -D$, \emph{always}. This effectively means that in \eqref{eq:Estar1}, $F^{*}_{\gamma, D}(\eps) = \emptyset$ and the only element in $E^*$ in \eqref{eq:Estar22} is $Z=-D$. Setting $z^* = \lfloor 0.5-\frac{\eps}{2\gamma}\rfloor = -D$ prevents it from  unnecessarily becoming too small, thereby avoiding a plateau. That is, we propose to choose $\gamma$ such that the first case in \eqref{eq:deltafinal1} always hold with equality $\lfloor 0.5-\frac{\eps}{2\gamma}\rfloor = -D$.\footnote{Note that we are not claiming this choice for $\gamma$ will minimise $\delta$ overall. This is because for simplicity of analysis, we are not considering both cases of \eqref{eq:deltafinal1} jointly to select the best $\gamma$ for a given $\eps$ and $D$. Our proposed method is a heuristic technique, which focuses on optimising the first case in \eqref{eq:deltafinal1} and obtains an analytical achievable expression for $\delta$ in terms of $\eps$ and $D$. It is intuitive that focusing on the first case of \eqref{eq:deltafinal1} should be a good choice, as it does not suffer from additional terms for $\delta$. See Fig. \ref{fig:deltaprop3} for a numerical corroboration.} This will give $\delta_{D}(\eps) = C(\eps)e^{-\gamma(\eps) D^2}$. We drop the dependence of $\delta$ on $\gamma$, as $\gamma$ will now be determined as a function of $\eps$ (for a fixed $D$).

Note that $\lfloor u \rfloor \leq u < \lfloor u \rfloor +1$. To have $z^* = \lfloor 0.5-\frac{\eps}{2\gamma}\rfloor = -D$, we need to ensure the following is satisfied
 \begin{align}\label{zstaropt21}
-D &\leq 0.5-\frac{\eps}{2\gamma} < -D+1 
\quad \Rightarrow \quad \frac{\eps}{2D+1} \leq \gamma < \frac{\eps}{2D-1}.
\end{align}
Therefore, the proposed range for $\gamma$ as a function of $\eps$ is as follows
 \begin{align}\label{zstaropt31}
\frac{\eps}{2D+1} \leq \gamma(\eps) < \frac{\eps}{2D-1}.
\end{align}
We now find what range for the variance of the TableBuilder is required to ensure the desired $\gamma(\eps)$. It turns out that we can find the corresponding range for $V(\eps)$ in analytical closed-form. Recall \eqref{eq:gammax1}, which is polynomial in $x = e^{-\gamma}$, but is affine in $V$. We can solve \eqref{eq:gammax1} for $V$ in terms of $x = e^{-\gamma}$:
\begin{align}\label{eq:Vsolution1}
V = \frac{\sum_{z=1}^D 2z^2e^{-\gamma z^2}}{2\sum_{z=1}^De^{-\gamma z^2} +1}.
\end{align}
It can be verified that $V$ in \eqref{eq:Vsolution1} is an increasing function of $x = e^{-\gamma}$ or a decreasing function of $\gamma$. Therefore, based on \eqref{zstaropt31} and \eqref{eq:Vsolution1}, the proposed range for $V(\eps)$ is
\begin{align}\label{eq:Vsolution2}
\frac{\sum_{z=1}^D 2 z^2e^{-\frac{\eps}{2D-1}z^2}}{2\sum_{z=1}^D e^{-\frac{\eps}{2D-1}z^2}+1} < V(\eps) \leq \frac{\sum_{z=1}^D 2 z^2e^{-\frac{\eps}{2D+1}z^2}}{2\sum_{z=1}^D e^{-\frac{\eps}{2D+1}z^2}+1}.
\end{align}
And from \eqref{eq:C1}, the desired range for $C$ (which is decreasing in $x=e^{-\gamma}$ or increasing in $\gamma$) is
 \begin{align}\label{eq:çsolution1}
 \frac{1}{2\sum_{z=1}^D e^{- \frac{\eps}{2D+1}z^2}+1} \leq C(\eps) < \frac{1}{2\sum_{z=1}^D e^{- \frac{\eps}{2D-1}z^2}+1}.
 \end{align}
Finally, we detail $\delta_D(\eps) = C(\eps)e^{-\gamma(\eps) D^2}$, which also has a range. It can be verified that $$\delta = Ce^{-\gamma D^2} = \frac{e^{-\gamma D^2}}{2\sum_{z=1}^D e^{-\gamma z^2}+1},$$ is an increasing function of $x = e^{-\gamma}$ or a decreasing function of $\gamma$. The obtained range for $\delta$ is
  \begin{align}\label{eq:deltafinal21}
\frac{e^{-\frac{\eps}{2D-1}D^2}}{2\sum_{z=1}^D e^{- \frac{\eps}{2D-1}z^2}+1} < \delta_D(\eps) \leq \frac{e^{-\frac{\eps}{2D+1}D^2}}{2\sum_{z=1}^D e^{- \frac{\eps}{2D+1}z^2}+1}.
\end{align}
Define $\iota$ satisfying $-\frac{\eps}{2D-1}+\frac{\eps}{2D+1} = \frac{-2\eps}{4D^2-1} \leq \iota  <0 $. As $\iota\to 0^-$, we can asymptotically set $\gamma(\eps) \to \left(\frac{\eps}{2D-1}\right)^-$, resulting in the following TableBuilder variance 
\begin{align}\label{eq:Vsolution2222}
V(\eps) \to \left(\frac{\sum_{z=1}^D 2 z^2e^{-\frac{\eps}{2D-1}z^2}}{2\sum_{z=1}^D e^{-\frac{\eps}{2D-1}z^2}+1}\right)^+,
\end{align}
to achieve 
\begin{align}\label{eq:delatasymp}
   \delta_D(\eps) \to \left(\frac{e^{-\frac{\eps}{2D-1}D^2}}{2\sum_{z=1}^D e^{- \frac{\eps}{2D-1}z^2}+1}\right)^+.
 \end{align}
 We summarise the results of this subsection in the following proposition. 
\begin{proposition}\label{prop:deltaopt}
Consider the mechanism in \eqref{eq:mechanism} for the single counting query $q$. For any given $\eps >0$, $D \in \N$ and $ 0< -\iota \leq \frac{2\eps}{4D^2-1}$,\footnote{Note, we use $-\iota$ in the formulae.} the TableBuilder mechanism with the following noise pmf
 \begin{align}\label{eq:pzopt}
p_Z(Z=z) = \frac{e^{-(\frac{\eps}{2D-1}-\iota)z^2}}{2\sum_{z=1}^D e^{- (\frac{\eps}{2D-1}-\iota)z^2}+1}, \quad z \in [-D,D],
\end{align}
and noise variance
\begin{align}\label{eq:Vsolution222}
V(\eps) = \frac{\sum_{z=1}^D 2 z^2e^{-(\frac{\eps}{2D-1}-\iota)z^2}}{2\sum_{z=1}^D e^{-(\frac{\eps}{2D-1}-\iota)z^2}+1},
\end{align}
will achieve $(\eps,\delta)$-DP such that
 \begin{align}\label{eq:deltafinalopt1}
\delta :=\delta_D(\eps) = \frac{e^{-(\frac{\eps}{2D-1}-\iota)D^2}}{2\sum_{z=1}^D e^{- (\frac{\eps}{2D-1}-\iota)z^2}+1}.
\end{align}
\end{proposition}

Fig. \ref{fig:2A} plots the analytical asymptotic expression for $\delta$ in \eqref{eq:delatasymp} versus $\eps$ for two noise support parameters $D = 11$ and $D=15$. The plateaus in Fig. \ref{fig:deltas} have disappeared and as $D$ increases,  $\delta$ decreases. For comparison, we also plot the best possible $\delta$, which is found numerically by varying $\gamma$ from $0.0001$ to $0.3$ in linear steps of $0.0001$, evaluating $\delta_{\gamma, D}(\eps)$ using \eqref{eq:deltafinal1}, and choosing the minimum $\delta$ possible. The gaps vary from being small to zero and corroborate our intuition that focusing on the first case of \eqref{eq:deltafinal1} and optimising it as described above is a good design strategy. Fig. \ref{fig:2B} plots the analytical expression for variance $V$ in \eqref{eq:Vsolution2222} versus $\eps$ for the corresponding two noise support parameters $D = 11$ and $D=15$.

\FigTwoTBNoiseSupport{$\delta$ versus $\eps$: analytical $\delta$ from \eqref{eq:delatasymp} versus numerically optimised values using \eqref{eq:deltafinal1}.}{Analytical TableBuilder variance from \eqref{eq:Vsolution2222} to achieve the corresponding analytical $\delta$ in Fig. \ref{fig:2A}.}{Plots of $\delta_{D}(\eps)$ and $V(\eps)$ versus $\eps$ for two TableBuilder noise support parameters $D=11$ and $D=15$.}

\subsection{A TableBuilder Noise Design Guide} \label{sec:designguide}

In some applications, it may be desirable to achieve a specific $(\eps, \delta)$-DP measure for the ABS perturbation methodology. In this subsection, we use the results in Subsection \ref{sec:optvar1} to prescribe a simple method for analytically choosing the parameters of the perturbation, that is, the support $D$ and the variance $V$ to achieve a desired $(\eps, \delta)$-DP. 

\begin{enumerate}
\item Start with the desired $\eps > 0$ and $1<\delta<1$ as inputs.
\item For the desired $\eps$, linearly increase the support $D = 1, 2, \cdots$ and evaluate $\delta_D(\eps)$ using \eqref{eq:deltafinalopt1} with $0 <-\iota \ll \frac{2\eps}{4D^2-1}$, until the desired $\delta$ (or the first value smaller than $\delta$) is reached. Select the last evaluated $D$, denoted by $D^*$, as the perturbation noise support parameter. Hence, $\Z = [-D^*,D^*]$.
\item The TableBuilder noise variance $V$ to support the desired $(\eps,\delta)$ is given by \eqref{eq:Vsolution222} using $\iota$ and the value for $D=D^*$ found in the previous step.
\item The TableBuilder noise pmf is given by 
\begin{align}\label{eq:optimalpmfmethod}
p_Z(Z=z) = \frac{e^{-(\frac{\eps}{2D^*-1}-\iota)z^2}}{2\sum_{z=1}^{D{^*}} e^{- (\frac{\eps}{2D^*-1}-\iota)z^2}+1}, \quad z \in [-D^*,D^*].
\end{align}
\end{enumerate}

We now demonstrate how this routine works via an example.

\begin{example}\label{ex:1}
Let us assume the desired privacy target is $\eps = 0.5$ and $\delta = 10^{-4}$. We find that the smallest $D$ that satisfies \eqref{eq:deltafinalopt1} with $-\iota =  \frac{2\eps}{10(4D^2-1)}$ is $D^* = 25$ resulting in $\delta \approx 9.91\times 10^{-5}$. The corresponding perturbation variance is $V \approx 49.00$ and $-\iota = \frac{2\times 0.5}{10(4\times 25^2-1)} \approx 4\times 10^{-5}$. So the overall perturbation pmf using \eqref{eq:optimalpmfmethod} is
\begin{align}\label{eq:optimalpmfexample}
p_Z(Z=z) = \frac{e^{-(\frac{0.5}{2\times25-1}-4\times 10^{-5})z^2}}{2\sum_{z=1}^{25} e^{- (\frac{0.5}{2\times 25-1}-4\times 10^{-5})z^2}+1}, \quad z \in [-25,25].
\end{align}
For example, if we evaluate the above pmf at $Z = 0$, $Z=\pm 1$, $Z= \pm 2$, $\cdots$, $Z = \pm 12$, $\cdots$, $Z = \pm 24$, and $Z = \pm 25$ we get
\begin{align}\label{eq:optimalpmfvalues}
\begin{matrix}
p_Z(Z=0) \approx 0.056895481243871,  \\ p_Z(Z=-1) = p_Z(Z=1) \approx0.056320120792644,  \\ p_Z(Z=-2) = p_Z(Z=2) \approx0.054628714970934, \\ \vdots\\ p_Z(Z=-12) = p_Z(Z=12) \approx 0.016632589297126, \\ \vdots\\p_Z(Z=-24) = p_Z(Z=24) \approx0.000163117271714, \\ p_Z(Z=-25) = p_Z(Z=25) \approx0.000099129808160.
   \end{matrix}
\end{align}
Note that $P(Z=-z) = P(Z=z)$. The plot of the pmf is shown in Fig.~\ref{fig:exampl1}.

\FigThreeExampleOne{Perturbation  distribution in Example 1.}

\end{example}

\section{CELL KEY METHODOLOGY} \label{sec:cellkey}

The ABS developed the cell key method to ensure that users cannot circumvent perturbation by making repeated requests for the same table. If the disclosure protection mechanism failed to deliver a consistent  random perturbation, then  a user  could  obtain different versions of the same table.  Comparing the cell values across these different versions might reveal some information about the original table. This risk is particularly important to address in the context of the ABS TableBuilder where there is no restriction to prevent a user requesting the same table many times \cite{Leaver2009}.

The cell key method assigns a pseudo-random number (also known as record key) to each record of the micro dataset. Record keys, $\Rkey_i$, are positive integers less than $2^{32}$. In \cite{thompson2013},  record keys of size $2^{32}$ were further processed (were combined byte-by-byte) to give cell keys of size $2^8$. But this low cell key size was mainly implemented to reduce the complexity of lookup tables for sampling from a quantised perturbation noise distribution. However, this small cell key size is not strictly necessary. As we will see in the next section, larger cell key sizes are needed to maintain desired DP measures. 

Therefore, in this section, we extend the cell key described in \cite{thompson2013} to allow cell keys to be a power of 2, which can be as high as $2^{32}$. The cell key size is denoted by $\CellKey$. When a table is constructed, the record keys are summed over each cell, to give
\begin{align}\label{eq:cellkey}
\CellVal ^j = \sum^{N}_{i=1} \Rkey^j_i(\text{ modulo } \bigN),
\end{align}
where the cell key has four components $j=1, \cdots, 4$ components and $\bigN$ is a large prime number and we take the modulo to prevent integer overflows when we sum the pseudo-random numbers. The final $\CellVal$ is determined  as follows
\begin{align}\label{eq:cellkeybitxor}
\CellVal  = \CellVal^1 \oplus \CellVal^2 \oplus \CellVal^3 \oplus \CellVal^4,
\end{align}where $\oplus$ is the bitwise XOR operator. The values $\CellVal^1$, $\CellVal^2$, $\CellVal^3$, $\CellVal^4$ are the four binary components derived from representing cell key as a binary number up to 32 bits. We will use this $\CellVal$ and its size $\CellKey$ in the next section for direct sampling from the perturbation noise. 

To summarise, we assume that $\CellVal$ values are uniformly generated in the range $[0,\CellKey-1]$, where $\CellKey$ is a power of 2. Typical values $2^8$, $2^{16}$, or $2^{32}$ will be studied here, but other values are also possible.

\section{SAMPLING AND ITS IMPACT ON $(\eps, \delta)$-DP} \label{sec:sampling}

\subsection{Sampling} For sampling, we first scale and quantise the cumulative mass function (cmf) of the proposed perturbation method in Proposition \ref{prop:deltaopt} according to the following procedure:

\begin{enumerate}
\item For a given $\epsilon$, $D$,  $\iota$, the pmf of ABS perturbation method, $p_Z$, is given by \eqref{eq:pzopt}. We first compute its cmf as 
\begin{align}\label{eq:czopt}
c_Z(Z=z) = \Prob[Z\leq z] = \sum_{z' = -\infty}^{z} p_Z(Z=z'), \quad z \in [-D,D],
\end{align}
where clearly $c_Z(Z=z) = 0$ for any $z < -D$ and $c_Z(Z=z) = 1$ for any $z \geq D$.
\item Then, given the maximum cell key size $\CellKey$, we scale and quantise  $c_Z$ into $c_Z^Q$ as follows:
\begin{align}\label{eq:czoptQ}
c_Z^Q(Z=z) = \lceil c_Z(Z=z)\times\CellKey \rceil,
\end{align}
where $\lceil\cdot\rceil$ is the integer ceiling function. This will ensure that the minimum and maximum bounds 0, and 1 in $c_Z$ will correspond to  0 and $\CellKey$ in $c_Z^Q$, respectively. 

\item The values of $c_Z^Q$ are stored in a lookup table of size $2D+1$. Since $D$ is usually small, this lookup table can be saved in a memory-efficient manner. 

\item When a $\CellVal$ is generated according to \eqref{eq:cellkeybitxor}, we use the lookup table to get a sample from the distribution as follows. For a given value $\CellVal$ in the range $[0, \CellKey-1]$, we output the sample $S$ as follows:
\begin{align}\label{eq:outputQ}
c_Z^Q(Z=z) \leq \CellVal < c_Z^Q(Z=z+1) \quad \Rightarrow \quad S = z+1.
\end{align}
\item If the cell key size, $\CellKey$ is small, it may happen that two or more consecutive $c_Z^Q$ may become identical. This means that some perturbation noise values $z$ can never be achieved. If this happens, $\CellKey$ must be increased or the parameters of the distribution must be adjusted to ensure the full support of the distribution can be achieved.  
\end{enumerate}
\begin{example}\label{ex:2}
Recall the pmf of in Example \ref{ex:1}.
Assume the cell key size is $\CellKey = 2^{32}$. We compute the scaled and quantised cmf $C_Z^Q$ according to \eqref{eq:czoptQ}. For example, $C_Z^Q$ values at $Z = -26, -25, -24, -23$, and $Z = 25$ are given as follows
\begin{align}\label{eq:optimalcmfvaluesQ}
\begin{matrix}
c_Z^Q(Z=-26) = 0,  \\ c_Z^Q(Z=-25) =  \lceil 0.000099129808160\times2^{32} \rceil = 425760,
\\ c_Z^Q(Z=-24)   = \lceil 0.0002622470798742925\times 2^{32} \rceil = 1126343,\\
c_Z^Q(Z=-23) =  \lceil 0.0005252540370388639\times 2^{32} \rceil = 2255949,\\
\vdots\\
c_Z^Q(Z=25) =  \lceil 1 \times 2^{32} \rceil = 2^{32}.
  \end{matrix}
\end{align}
The values of $c_Z^Q$ will be stored in a lookup table of size $2D+1 = 51$. Now imagine that $\CellVal = 2552$ is given according to \eqref{eq:cellkeybitxor}. Since $c_Z^Q(Z=-26) \leq 2552 < c_Z^Q(Z=-25) = 425760$, we output $S = -25$ as the ABS perturbation noise. As another example, assume $\CellVal = 1200124$ is given. Since $c_Z^Q(Z=-24) \leq 1200124 < c_Z^Q(Z=-23) = 2255949$, we should output $S = -23$ as the ABS perturbation noise, and so on.

Now assume that $\CellKey = 2^{8}$ is given instead. We can see that $c_Z^Q(Z=-25) = c_Z^Q(Z=-24) = c_Z^Q(Z=-23) = 1$. This means not all values in the support $[-D,D]$ can be realised in practice. Hence, we conclude that $\CellKey = 2^{8}$ is not a sufficient cell key size for this perturbation distribution. 
\end{example}

\subsection{Evaluating Post-Sampling Utility and Privacy Measures} It now remains to verify the properties of the scaled and quantised distribution in terms of bias, variance and $(\eps^Q, \delta^Q)$-DP, where the superscript $Q$ signifies values post sampling. To this end, we follow the procedures below.
\begin{enumerate}
\item We convert the scaled and quantised cmf $c_Z^Q$ in \eqref{eq:czoptQ} into the scaled and quantised pmf $p_Z^Q$ as follows:
\begin{align}\label{eq:pzoptQ}
p_Z^Q(Z=z) = \frac{c_Z^Q(Z=z) - c_Z^Q(Z=z-1)}{\CellKey}, \quad z \in [-D:D].
\end{align}
Note that we assume $\CellKey$ is chosen sufficiently large to ensure $p_Z^Q$ has full support over $[-D,D]$. In steps below, we use the shorthand $p^Q(z):=p_Z^Q(Z=z)$.
\item The resulting bias and variance of $p_Z^Q$ are computed as
\begin{align}\label{eq:bias:varQ}
B^Q &:= \sum_{z = -D}^{D} z p^Q(z),\\\label{eq:bias:varQ2}
V^Q &:= \sum_{z = -D}^{D} z^2 p^Q(z) - (B^Q)^2.
\end{align}
These metrics clearly depend on the cell key size, $\CellKey$. It should be intuitively understood the larger the $\CellKey$, the finer the quantisation will be and the closer the bias and variance of $p_Z^Q$ should be to its original continuous version.
\end{enumerate}

To understand the effective $(\eps^Q, \delta^Q)$-DP metric as a result of scaling and quantisation in $p_Z^Q$, we propose the following method.

\begin{enumerate}

\item Recalling the definition \eqref{eq:deltazJ}, we know that $z = -D$ always belongs to $E^*$. We then find the smallest $\eps^Q$ that ensures the ratio $\frac{p^Q(z)}{p^Q(z-1)}\leq e^{\eps^Q}$ is maintained for all other support values $z \in [-D+1:D]$. That is, the goal is to ensure no other mass in the support contributes to $\delta^Q$. In other words, we find the smallest effective $\eps^Q$ that ensures $E^* = \{-D\}$ remains as before. Therefore, we first define and compute $\eps^Q$ as follows:
\begin{align}\label{eq:epzQ}
\eps^Q = \arg\min\{\eps: \frac{p^Q(z)}{p^Q(z-1)}<e^\eps, z \in [-D+1,D]\}.
\end{align}
\item Once the effective $\eps^Q$ is obtained as above, the effective  $\delta^Q$ will be the maximum of the pmf $p_Z^Q$ at the two extreme support values and is given by
\begin{align}\label{eq:deltazQ}
\delta^Q = \max\{p^Q(-D), p^Q(D)\}.
\end{align}
Again, the cell key size, $\CellKey$ will play a main role on the resulting $(\eps^Q, \delta^Q)$ metric. The larger the $\CellKey$, the closer $(\eps^Q, \delta^Q)$ can get to the original $(\eps, \delta)$ metrics for the continuous case. Also, $\frac{1}{\CellKey}$ will pose a lower bound on how small $\delta^Q$ can get, as this is the smallest value that $P_Z^Q(Z=-D)$ or $P_Z^Q(Z=D)$ can have. 
\end{enumerate}

\begin{example}
Continuing on Example \ref{ex:1} and Example \ref{ex:2}, we can convert the scaled and quantised cmf $C_Z^Q$ according to \eqref{eq:pzoptQ} back to quantised pmf $P_Z^Q$. For example, at $Z = -25, -24, -23$, and $Z = 25$ we will have
\begin{align}\label{eq:optimalpmfvaluesQ}
\begin{matrix}
P_Z^Q(Z=-25) = \frac{425760}{2^{32}} \approx  0.000099129974842,
\\ P_Z^Q(Z=-24) = \frac{1126343-425760}{2^{32}} \approx 0.000163117190823,\\
P_Z^Q(Z=-23) = \frac{2255949-1126343}{2^{32}} \approx 0.0002630068920552731,\\
\vdots\\
P_Z^Q(Z=25) =  \frac{2^{32}-4294541537}{2^{32}} = 0.00009912974201142788.
  \end{matrix}
\end{align}
Note that the quantised pmf has lost its complete symmetry, compared to the original pmf in Example \ref{ex:2}. Its bias can be calculated from \eqref{eq:bias:varQ} to be $B^Q = -5.820766091346741\times 10^{-9}$. Its variance can be calculated from \eqref{eq:bias:varQ2} to be $V^Q = 49.002167175291106$, which are very close to the original zero-bias and design variance, respectively. 

Now, we compute $\eps^Q$ according to \eqref{eq:epzQ}, which gives $\eps^Q \approx 0.498037038323823$. Interestingly, this is slightly smaller than the design target $\eps = 0.5$. This is not unusual, since the quantisation is a non-linear operation and $\eps^Q$ can be lower or higher than $\eps$. We will investigate this further in the upcoming experiments. Finally, $\delta^Q = P_Z^Q(-D) \approx 9.9129974842\times 10^{-5}$ is computed according to \eqref{eq:deltazQ}, which is only slightly larger than the original $\delta$ in Example \ref{ex:1}.

\end{example}

\subsection{Experiments}
To study the effect of $\CellKey$ on the perturbation bias, variance and DP measures more systematically, we consider the following scenario. We set $D=10$, vary $\eps \in [0.1,2.5]$ in 0.1 steps, let $-\iota =  \frac{2\eps}{10(4D^2-1)}$ and follow the proposed quantisation procedure we described in the previous subsections. 

First, we find that bias $B^Q \approx -2.3\times 10^{-9}$ is lowest when $\CellKey= 2^{32}$. This deteriorates to $B^Q \approx -1.5\times 10^{-4}$ when $\CellKey= 2^{16}$ and to $B^Q \approx -0.04$ when $\CellKey= 2^{8}$. This confirms that $\CellKey$ has a clear effect on the post-sampled perturbation measures. Furthermore, for  $\CellKey= 2^{8}$ and $\CellKey= 2^{16}$ not all values of $\eps$ result in distributions with full support. 

Next, we define the normalised error in variance after quantisation as $$\frac{V^Q-V}{V}.$$ This normalised variance error is in the order of $10^{-11}$ and $10^{-10}$ for $\CellKey= 2^{32}$ and $\CellKey= 2^{16}$, respectively. However, when $\CellKey= 2^{8}$  the normalised variance error can be as high as $0.01$.

\begin{figure*}[!tbh]
		\centering
		\includegraphics[width=8.8cm]{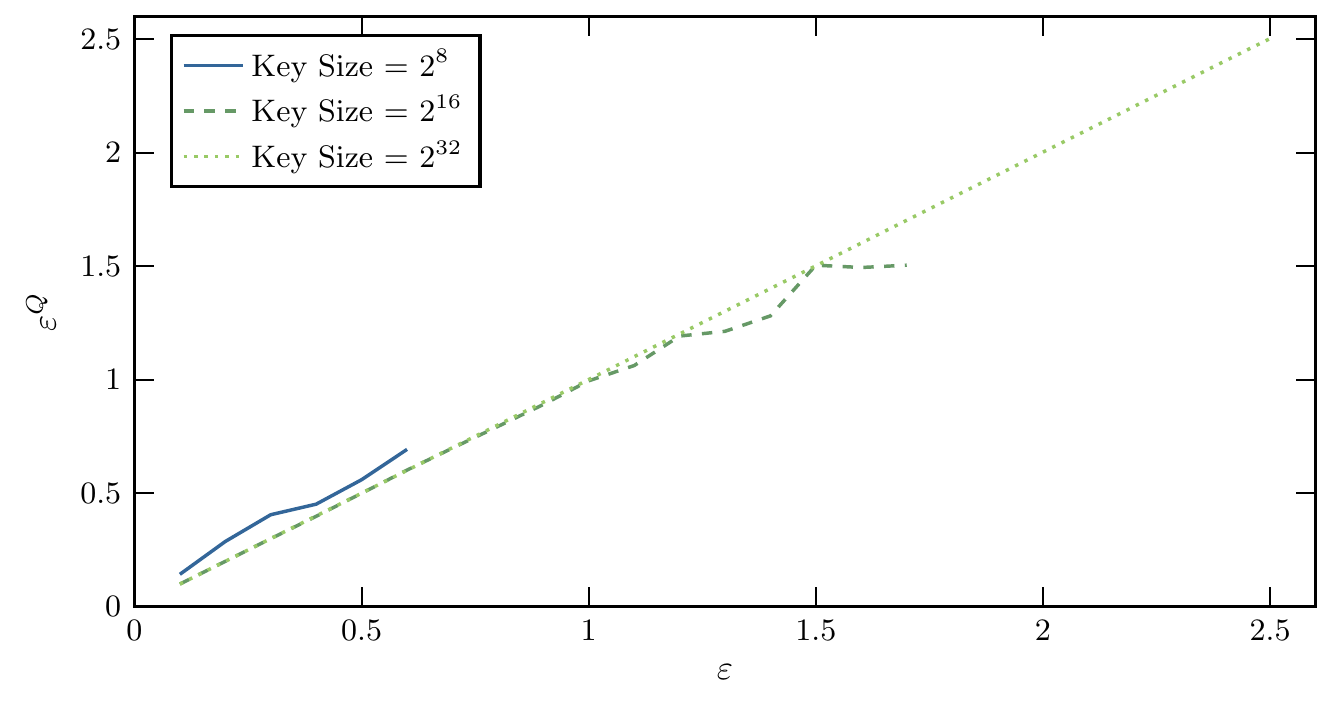}
		\caption{$\eps^Q$ versus $\eps$ for three values of cell key size. Overall, only $\CellKey = 2^{32}$ can closely follow the original $\eps$ across its entire range.}\label{fig:epsQ}
\end{figure*}

Fig.~\ref{fig:epsQ} shows the relation between $\eps$ at the time of design and the resulting $\eps^Q$ post sampling for three different values of $\CellKey$. $\eps \geq \eps^Q$ is desirable and $\eps^Q>\eps$ is not desirable. We see that when  $\CellKey = 2^8$, $\eps^Q>\eps$. When $\CellKey = 2^{16}$, $\eps^Q$ is either close to $\eps$ or slightly lower. The nonlinear/jittery behaviour is not unusual and is due to the nonlinear sampling scheme, which involves the integer ceiling function. However, the problem with both $\CellKey = 2^8$ and $\CellKey = 2^{16}$ values is that the quantised perturbation $P_Z^Q$ cannot provide full support due to the nonlinear quantisation and insufficiently large $\CellKey$. This is not acceptable since the designed support of $[-10,10]$ cannot be maintained which is the original design criterion. 
For $\CellKey = 2^8$, this happens after $\eps = 0.6$ and for $\CellKey = 2^{16}$, this happens after $\eps = 1.7$. Whereas, when $\CellKey = 2^{32}$, we see that $\eps^Q\approx \eps$ as  desired and the full support is maintained for all $\eps$ values under consideration.

\begin{figure*}[!tbh]
		\centering
		\includegraphics[width=8.8cm]{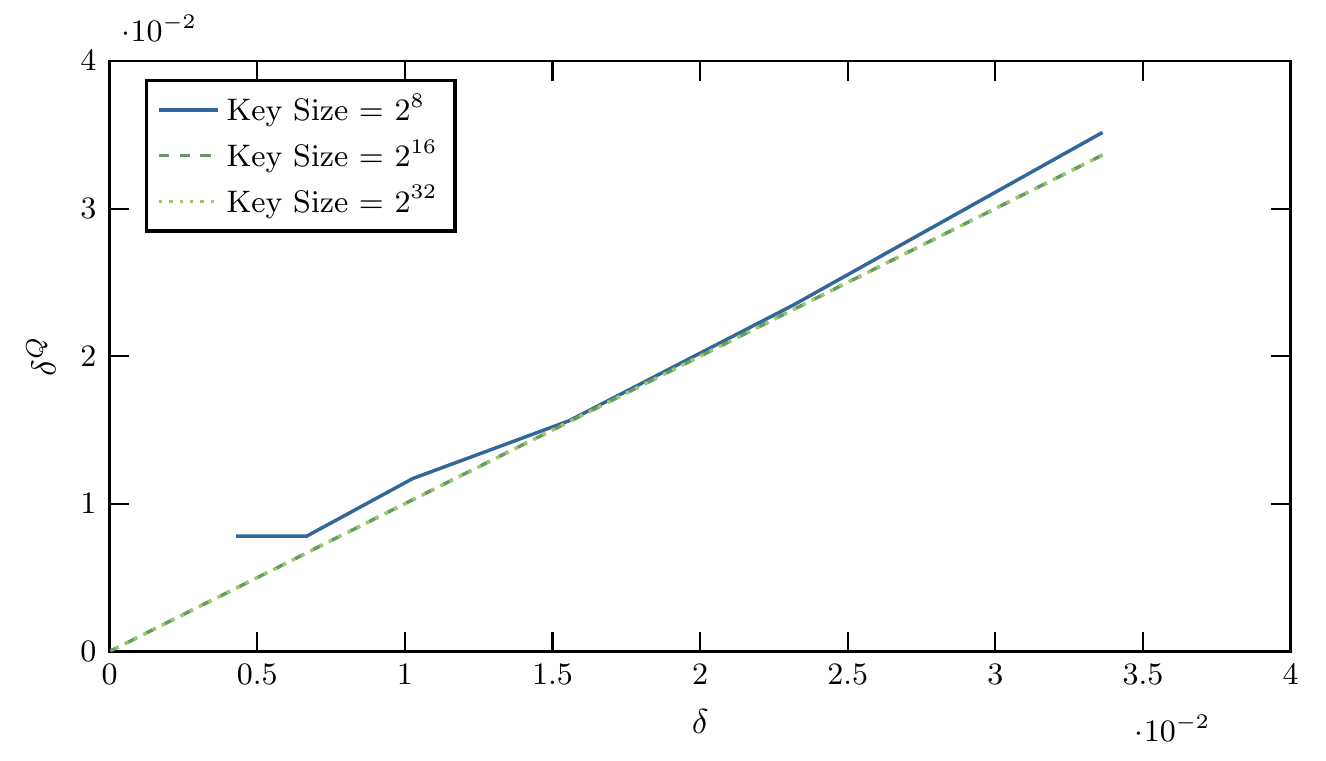}
		\caption{$\delta^Q$ versus $\delta$ for three values of cell key size. Overall, only $\CellKey = 2^{32}$ can closely follow the original $\delta$ across the entire range of $\eps$.}\label{fig:deltaQ}
\end{figure*}

Fig.~\ref{fig:deltaQ} shows the relation between $\delta$ at the time of design and the resulting $\delta^Q$ post sampling for three different values of $\CellKey$. $\delta \geq \delta^Q$ is desirable and $\delta^Q>\delta$ is not desirable. We see that when  $\CellKey = 2^8$, $\delta^Q>\delta$. When $\CellKey = 2^{16}$ or $\CellKey = 2^{32}$, $\delta^Q$ is almost identical to the original $\delta$. However, as mentioned before the caveat to using $\CellKey = 2^8$ or $\CellKey = 2^{16}$ is that the full support of perturbation noise  maintained cannot be maintained for all $\eps$ values under consideration.

\section{CONCLUSION} \label{sec:conclusion}

DP framework provides an opportunity to better quantify the confidentiality protection and data utility of the ABS perturbation methodology. We have proposed an alternative entropy maximisation approach which incorporates ($\eps$, $\delta$)-DP parameters for symmetric support. 

We have proposed an approach to expand the cell key row index size. We have shown the importance of having a larger cell key size to achieve the desired ($\eps$, $\delta$)-DP parameters in our quantised sampling approach. 

There are several potential areas of future research including (1) extending the method to consider asymmetrical perturbation distributions, (2) developing a framework to consider $\eps$ and $\delta$ parameters for dynamic table environments and (3) evaluating the performance against different types of perturbation distributions.

\bibliographystyle{IEEEtran}
\bibliography{DPABS}

\end{document}

%% file: preamble_arxiv.tex
\usepackage[ruled]{algorithm2e}
\usepackage{graphicx}
\usepackage{placeins} %FloatBarrier
\graphicspath{{./plots/}} 
\input{macros_arxiv.tex}

\usepackage{subcaption}
\usepackage{amsfonts}
\usepackage{amssymb, amsthm, amsmath}
\usepackage{cite}
\usepackage{tcolorbox}
\usepackage{url}
\def\eps{{\varepsilon}}

\newtheorem{example}{Example}%[section]
\usepackage[margin=1in]{geometry}
\newenvironment{enumerate*}%
{\begin{enumerate}%
		\setlength{\itemsep}{0pt}%
		\setlength{\parskip}{0pt}
		\vspace*{-\topsep}
	}%
	{\end{enumerate}
	\vspace*{-1.5ex}}

\newenvironment{itemize*}%
{\vspace*{0.1ex}
	\begin{itemize}%
		\setlength{\itemsep}{0pt}%
		\setlength{\parskip}{0pt}
		\vspace*{-\topsep}
	}%
	{\end{itemize}
	\vspace*{-1.5ex}
}

\newtheorem{definition}{Definition}

\newtheorem{proposition}{Proposition}

\theoremstyle{definition}

\newcommand{\N}{\mathbb{N}}
\newcommand{\X}{\mathcal{X}}
\newcommand{\Y}{\mathcal{Y}}
\newcommand{\Z}{\mathcal{Z}}
\newcommand{\E}{\mathbb{E}}
\newcommand{\Prob}{\mathbb{P}}
\newcommand{\CellKey}{\mathtt{KEYSIZE}}
\newcommand{\Rkey}{\mathtt{Rkey}}
\newcommand{\bigN}{\mathtt{bigN}}
\newcommand{\CellVal}{\mathtt{CellKey}}

%% file: macros_arxiv.tex
\newcommand{\FigOneTBNoise}[3]
{%\noindent
	\FloatBarrier
\begin{figure*}[!t]
	\begin{subfigure}[t]{0.48\textwidth}
		\centering
		{\includegraphics[width=1\columnwidth]{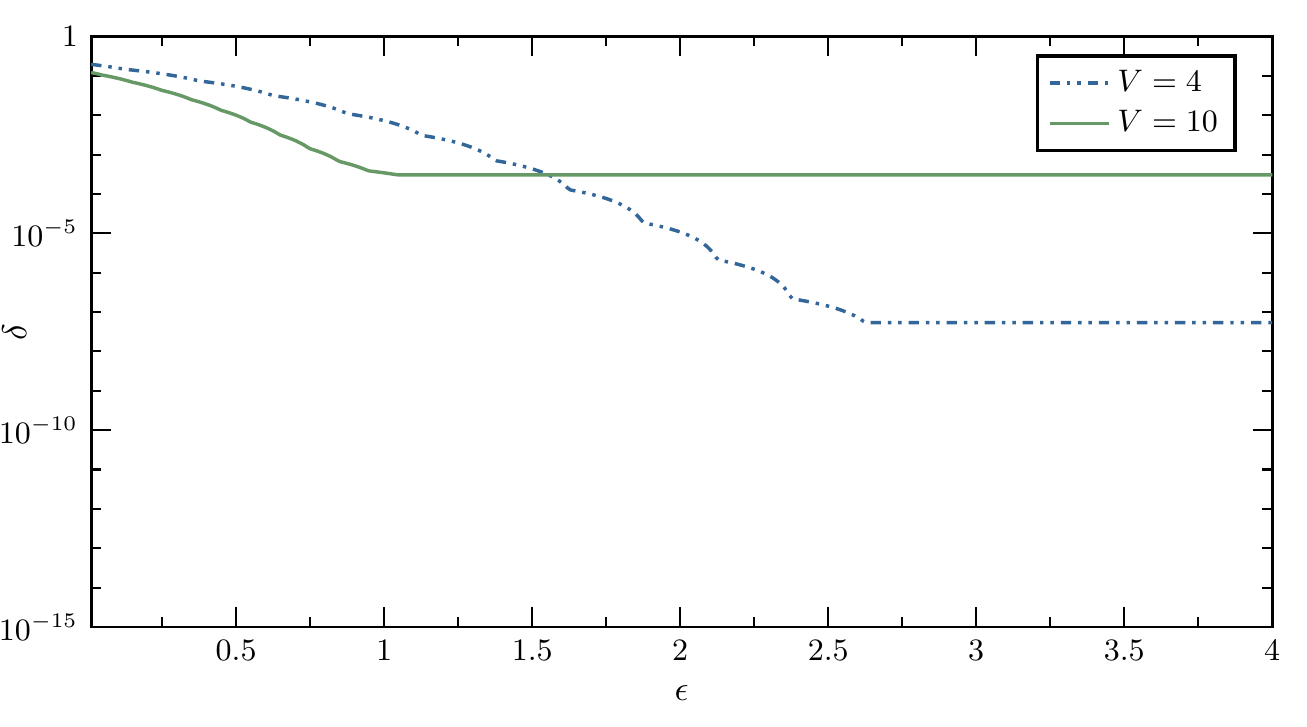}}
		\caption{#1} \label{fig:deltasA}   
	\end{subfigure}\hspace{4mm}
	\begin{subfigure}[t]{0.48\textwidth}
		\centering
		\includegraphics[width=1\columnwidth]{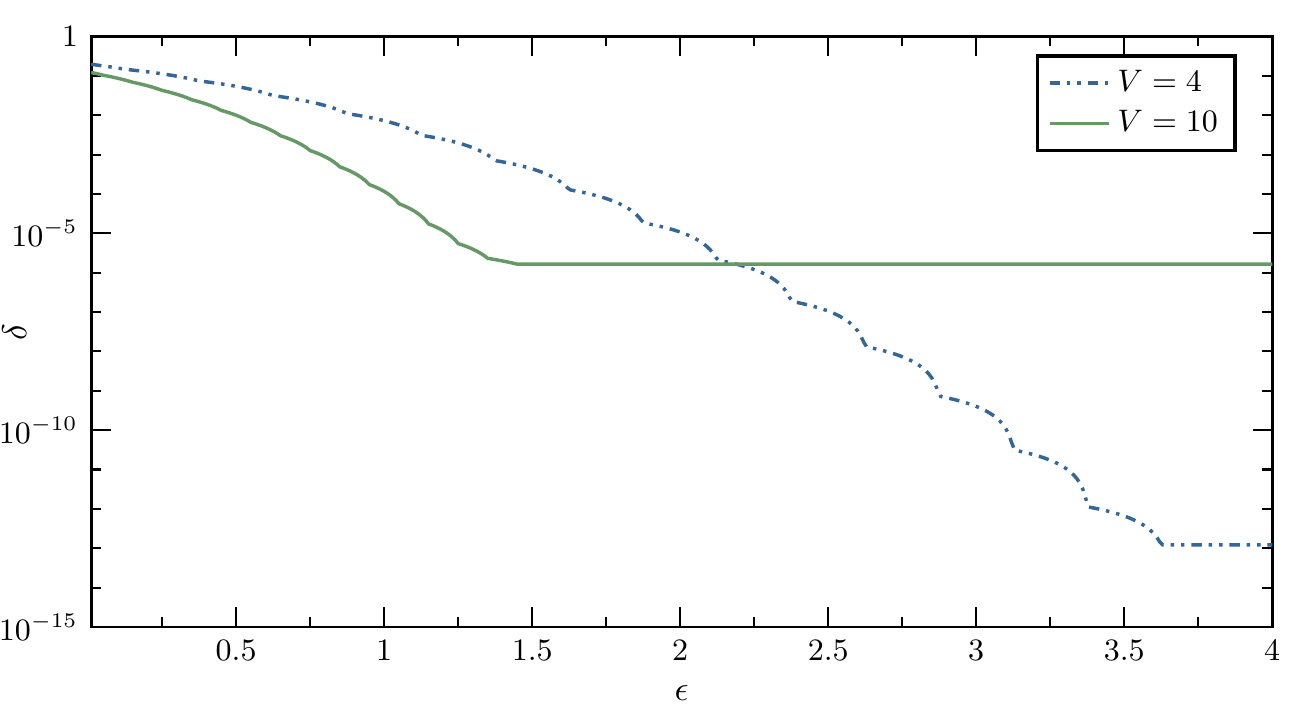}
		\caption{#2}\label{fig:deltasB}
	\end{subfigure}
	\caption{#3}
	\label{fig:deltas}
\end{figure*}
	\FloatBarrier
}

\newcommand{\FigTwoTBNoiseSupport}[3]
{\noindent
	\FloatBarrier
\begin{figure*}[!t]\begin{subfigure}[t]{0.45\textwidth}
    \centering
             \raisebox{0.40 em}{\includegraphics[width=1\columnwidth]{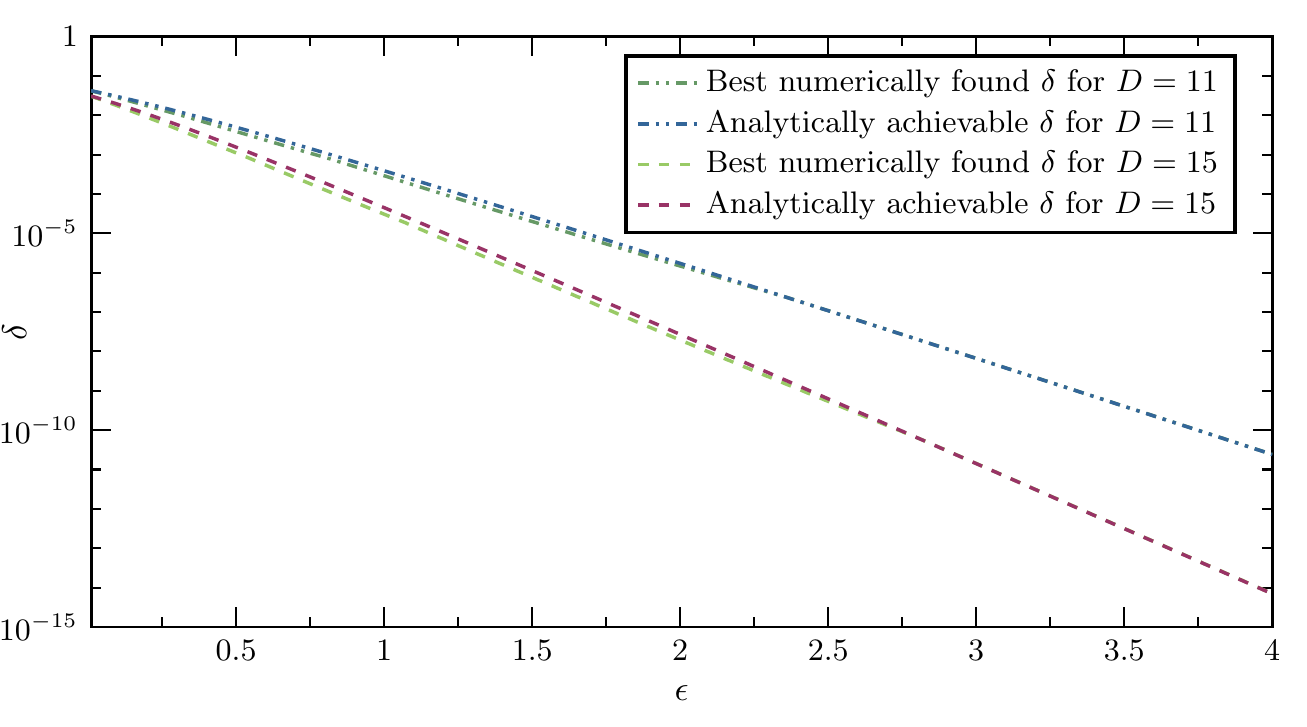}}
            \caption{#1}\label{fig:2A} \end{subfigure}\hspace{4mm}
            \begin{subfigure}[t]{0.47\textwidth}
    \centering
             \includegraphics[width=1\columnwidth]{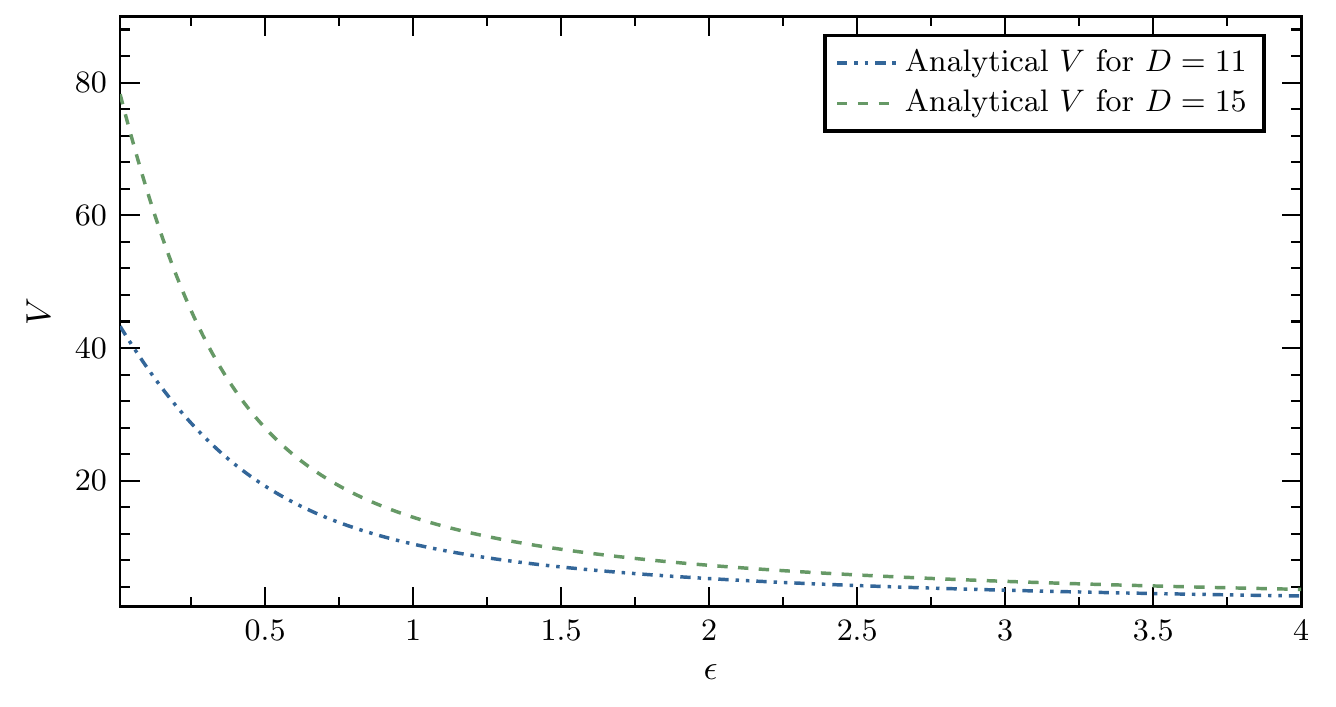}
            \caption{#2}\label{fig:2B} \end{subfigure}
   \caption{#3}
\label{fig:deltaprop3}
\end{figure*}
	\FloatBarrier
}

\newcommand{\FigThreeExampleOne}[1]
{\noindent
	\FloatBarrier
\begin{figure*}[!tbh]
    \centering
             \includegraphics[width=0.6\columnwidth]{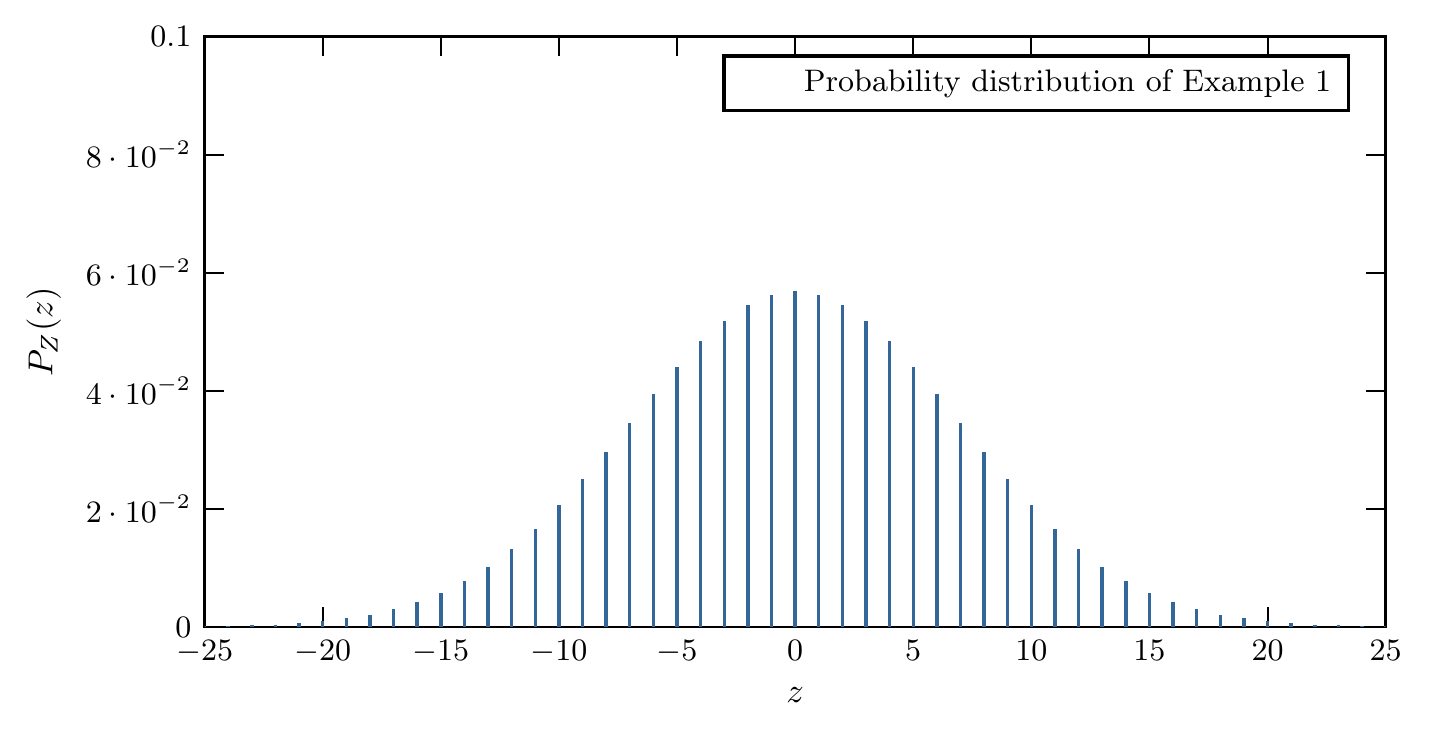}
            \caption{#1}\label{fig:exampl1}
\end{figure*}
	\FloatBarrier
}

%% file: ARXIV.bbl
\begin{thebibliography}{1}
\providecommand{\url}[1]{#1}
\csname url@samestyle\endcsname
\providecommand{\newblock}{\relax}
\providecommand{\bibinfo}[2]{#2}
\providecommand{\BIBentrySTDinterwordspacing}{\spaceskip=0pt\relax}
\providecommand{\BIBentryALTinterwordstretchfactor}{4}
\providecommand{\BIBentryALTinterwordspacing}{\spaceskip=\fontdimen2\font plus
\BIBentryALTinterwordstretchfactor\fontdimen3\font minus
  \fontdimen4\font\relax}
\providecommand{\BIBforeignlanguage}[2]{{%
\expandafter\ifx\csname l@#1\endcsname\relax
\typeout{** WARNING: IEEEtran.bst: No hyphenation pattern has been}%
\typeout{** loaded for the language `#1'. Using the pattern for}%
\typeout{** the default language instead.}%
\else
\language=\csname l@#1\endcsname
\fi
#2}}
\providecommand{\BIBdecl}{\relax}
\BIBdecl

\bibitem{ABS2022}
ABS, ``1005.0 - abs corporate plan, 2021-22,''
  \url{https://www.abs.gov.au/ausstats/abs@.nsf/Lookup/by\%20Subject/1005.0~2021-22~Main\%20Features~Objectives~6\#Objective3},
  2022, accessed: 2022-05-01.

\bibitem{Fraser2005}
B.~Fraser and J.~Wooton, ``A proposed method for confidentialising tabular
  output to protect against differencing,'' \emph{Monographs of Official
  Statistics: Work Session on Statistical Data Confidentiality}, pp. 299--302,
  2005.

\bibitem{Marley2011}
J.~K. Marley and V.~L. Leaver, ``A method for confidentialising user-defined
  tables: statistical properties and a risk-utility analysis,'' in
  \emph{Proceedings of the 58th Congress of the International Statistical
  Institute, ISI}, 2011, pp. 21--26.

\bibitem{Bailie2019}
J.~Bailie and C.-H. Chien, ``Abs perturbation methodology through the lens of
  differential privacy,'' \emph{Joint UNECE/Eurostat Work Session on
  Statistical Data Confidentiality, The Hague, Netherlands}, 2019.

\bibitem{dwork2006calibrating}
C.~Dwork, F.~McSherry, K.~Nissim, and A.~Smith, ``Calibrating noise to
  sensitivity in private data analysis,'' in \emph{Theory of cryptography
  conference}.\hskip 1em plus 0.5em minus 0.4em\relax Springer, 2006, pp.
  265--284.

\bibitem{dwork2014algorithmic}
C.~Dwork, A.~Roth \emph{et~al.}, ``The algorithmic foundations of differential
  privacy,'' \emph{Foundations and Trends{\textregistered} in Theoretical
  Computer Science}, vol.~9, no. 3--4, pp. 211--407, 2014.

\bibitem{balle18a}
B.~Balle and Y.-X. Wang, ``Improving the {G}aussian mechanism for differential
  privacy: Analytical calibration and optimal denoising,'' in \emph{Proceedings
  of the 35th International Conference on Machine Learning}, ser. Proceedings
  of Machine Learning Research, J.~Dy and A.~Krause, Eds., vol.~80.\hskip 1em
  plus 0.5em minus 0.4em\relax PMLR, 10--15 Jul 2018, pp. 394--403.

\bibitem{Leaver2009}
V.~Leaver, ``Implementing a method for automatically protecting user-defined
  census tables,'' \emph{Joint ECE/Eurostat Worksession on Statistical
  Confidentiality in Bilbao (December 2009), http://www. unece.
  org/stats/documents/2009.12. confidentiality. htm}, 2009.

\bibitem{thompson2013}
G.~Thompson, S.~Broadfoot, and D.~Elazar, ``Methodology for the automatic
  confidentialisation of statistical outputs from remote servers at the
  australian bureau of statistics,'' \emph{Joint UNECE/Eurostat work session on
  statistical data confidentiality}, pp. 28--30, 2013.

\end{thebibliography}
